\documentstyle[amssymb,prb,preprint,aps]{revtex}

\draft

\begin{document}
\title{Specific heat and magnetic measurements in Nd$_{0.5}$Sr$_{0.5}$MnO$_{3}$ and
R$_{0.5}$Ca$_{0.5}$MnO$_{3}$ (R=Nd, Sm, Dy and Ho) samples}
\author{J. L\'{o}pez and O. F. de Lima}
\address{Instituto de F\'{i}sica Gleb Wataghin, Universidade Estadual de Campinas,\\
UNICAMP, 13083-970, Campinas, SP, Brazil}
\maketitle

\begin{abstract}
We have made a magnetic characterization of Nd$_{0.5}$Sr$_{0.5}$MnO$_{3}$, Nd%
$_{0.5}$Ca$_{0.5}$MnO$_{3}$, Sm$_{0.5}$Ca$_{0.5}$MnO$_{3}$, Dy$_{0.5}$Ca$%
_{0.5}$MnO$_{3}$ and Ho$_{0.5}$Ca$_{0.5}$MnO$_{3}$ polycrystalline samples.
Ferromagnetic, antiferromagnetic and charge ordering transitions in our
samples agree with previous reports. We also studied specific heat
measurements with applied magnetic fields between 0 and 9 T and temperatures
between 2 and 300 K in all cases. Each curve was successfully fitted at high
temperatures by an Einstein model with three optical phonon modes. Close to
the charge ordering and ferromagnetic transition temperatures the specific
heat curves showed peaks superposed to the characteristic response of the
lattice oscillations. The entropy variation corresponding to the charge
ordering transition was higher than the one corresponding to the
ferromagnetic transition. The external magnetic field seems to have no
effect in specific heat of the CO phase transition.
\end{abstract}


\section{Introduction}

The physical properties in charge ordering (CO) manganese perovskites are
thought to arise from the strong competition involving a ferromagnetic
double exchange interaction, an antiferromagnetic superexchange interaction,
and the spin-phonon coupling\cite{Radaelli}$^{,}$\cite{Zhao2}$^{,}$\cite
{Moritomo}$^{,}$\cite{Xiao}$^{,}$\cite{Mori}$^{,}$\cite{J.López1}$^{,}$\cite
{J.López2}. These interactions are determined by intrinsic parameters such
as doping level, average cationic size, cationic disorder and oxygen
stoichiometry. CO compounds are particularly interesting because spin,
charge and orbital degrees of freedom are at play simultaneously and
classical simplifications, that neglect some of these interactions, do not
work. More detailed information on the physics of manganites can be found in
a review paper by Myron B. Salamon and Marcelo Jaime\cite{Myron}.

L. Ghivelder et al.\cite{Ghivelder} reported specific heat measurements in
LaMnO$_{3+\delta }$ samples and found that the specific heat at low
temperature is very sensitive to small variations of $\delta $, similarly to
results published by W. Schnelle et al.\cite{Schnelle} in a Nd$_{0.67}$Sr$%
_{0.33}$MnO$_{3-\delta }$ sample. In this latter work, and also in a paper
by J. E. Gordon et al.\cite{Gordon}, a Schottky-like anomaly was found at
low temperatures. They associated this result to the magnetic ordering of Nd$%
^{3+}$ ions and to the crystal-field splitting. F. Bartolom\'{e} et al.\cite
{Bartolomé} also found a Schottky-like anomaly in a closely related compound
of NdCrO$_{3}$. They proposed a crystal-field energy level scheme in
agreement with neutron-scattering studies in the same sample. In two papers
V. N. Smolyaninova et al.\cite{Smolyaninova1}$^{,}$\cite{Smolyaninova2}
studied the low temperature specific heat in Pr$_{1-x}$Ca$_{x}$MnO$_{3}$ (0.3%
\mbox{$<$}%
x%
\mbox{$<$}%
0.5) and La$_{1-x}$Ca$_{x}$MnO$_{3}$ (x=0.47, 0.5 and 0.53). They fitted the
data with an excess specific heat, C%
\'{}%
(T), of non-magnetic origin associated with charge ordering. They also
showed that a magnetic field, sufficiently high to induce a transition from
the charge ordered state to the ferromagnetic metallic state, did not
completely remove C%
\'{}%
(T). However, no Schottky anomaly was found in any of these compounds.

We have already presented specific heat measurements with applied magnetic
fields between 0 and 9 T and temperatures between 2 and 30 K for Nd$_{0.5}$Sr%
$_{0.5}$MnO$_{3}$, Nd$_{0.5}$Ca$_{0.5}$MnO$_{3}$, Sm$_{0.5}$Ca$_{0.5}$MnO$%
_{3}$, Dy$_{0.5}$Ca$_{0.5}$MnO$_{3}$ and Ho$_{0.5}$Ca$_{0.5}$MnO$_{3}$
samples. All these compounds presented a Schottky-like anomaly at low
temperatures\cite{JLópez3}$^{,}$\cite{JLópez4}. Here, we report a general
magnetic characterization and specific heat measurements in the full
temperatures interval, between 2 and 300 K, for the same five samples. As
far as we know, the high temperature interval of the specific heat
measurements in these compounds have not been published yet.

\section{Experimental methods}

Polycrystalline samples of Nd$_{0.5}$Sr$_{0.5}$MnO$_{3}$, Nd$_{0.5}$Ca$%
_{0.5} $MnO$_{3}$ and Ho$_{0.5}$Ca$_{0.5}$MnO$_{3}$ were prepared by the
sol-gel method\cite{Paulo}. Stoichiometric parts of Nd$_{2}$O$_{3}$ (Ho$_{2}$%
O$_{3}$) and MnCO$_{3}$ were dissolved in HNO$_{3}$ and mixed to an aqueous
citric acid solution, to which SrCO$_{3}$ or CaCO$_{3}$\ was added. The
mixed metallic citrate solution presented the ratio citric acid/metal of 1/3
(in molar basis). Ethylene glycol was added to this solution, to obtain a
citric acid/ethylene glycol ratio 60/40 (mass ratio). The resulting solution
was neutralized to pH$\sim $7 with ethylenediamine. This solution was turned
into a gel, and subsequently decomposed to a solid by heating at 400 $^{o}$%
C. The resulting powder was heat-treated in vacuum at 900 $^{o}$C for 24
hours, with several intermediary grindings, in order to prevent formation of
impurity phases. This powder was pressed into pellets and sintered in air at
1050 $^{o}$C for 12 hours.

Polycrystalline samples of Sm$_{0.5}$Ca$_{0.5}$MnO$_{3}$ and Dy$_{0.5}$Ca$%
_{0.5}$MnO$_{3}$ were prepared from stoichiometric amounts of Sm$_{2}$O$_{3}$
or Dy$_{2}$O$_{3}$, CaO, and MnO$_{2}$ by standard solid-state reaction
method. All the powders were mixed and ground for a long time in order to
produce a homogeneous mixture. First, the mixture was heated at 927 $^{o}$C
for 24 hours and after that it was ground and heated at 1327 $^{o}$C (72
hours) and 1527 $^{o}$C (48 hours). X-ray diffraction measurements indicated
high quality samples in all cases.

The magnetization measurements were done with a Quantum Design MPMS-5S SQUID
magnetometer. Specific heat measurements were done with a Quantum Design
PPMS calorimeter. The PPMS used the two relaxation time technique, and data
was always collected during sample cooling. The intensity of the heat pulse
was calculated to produce a variation in the temperature bath between 0.5 \%
(at low temperatures) and 2\% (at high temperatures). Experimental errors
during the specific heat and magnetization measurements were lower than 1 \%
for all temperatures and samples.

\section{Results and Discussion}

\subsection{Magnetization measurements}

Figure 1 shows the magnetization temperature dependence, measured with a 5 T
applied magnetic field in field cooling conditions, in polycrystalline
samples of Nd$_{0.5}$Sr$_{0.5}$MnO$_{3}$, Nd$_{0.5}$Ca$_{0.5}$MnO$_{3}$, Sm$%
_{0.5}$Ca$_{0.5}$MnO$_{3}$, Dy$_{0.5}$Ca$_{0.5}$MnO$_{3}$ and Ho$_{0.5}$Ca$%
_{0.5}$MnO$_{3}$. The curves are plotted with a logarithmic scale in the
y-axes to facilitate comparisons. Charge ordering transition temperatures (T$%
_{CO}$) are indicated by arrows at 160, 250, 270, 280 and 271 K,
respectively. These temperatures are associated to peaks in the
magnetization curves, in agreement with previous reports \cite{Kajimoto}$^{,}
$ \cite{Millange}$^{,}$\cite{Tokura}$^{,}$\cite{Terai}. It is interesting to
note that the relation between the charge ordering temperature and the
antiferromagnetic ordering temperature (T$_{N}$) changes from one sample to
the other\cite{Kajimoto}$^{,}$ \cite{Millange}$^{,}$\cite{Tokura}$^{,}$\cite
{Terai}. In the first case they are approximately coincident, in the second
and third cases the charge ordering temperatures are much higher, and in the
fourth and fifth cases a long range antiferromagnetic transition is not
observed.

The Nd$_{0.5}$Sr$_{0.5}$MnO$_{3}$ sample presented a ferromagnetic
transition at T$_{C}\thickapprox $244 K and an antiferromagnetic transition
at T$_{N}\thickapprox $160 K. The Nd$_{0.5}$Ca$_{0.5}$MnO$_{3}$ compound
presented a strong magnetization maximum near T$_{CO}$, but showed an
unexpected minimum close to the antiferromagnetic transition temperature T$%
_{N}\thickapprox $160 K. Usually an antiferromagnetic transition is
accompanied by a maximum in the temperature dependence of magnetization. The
antiferromagnetic transition in Sm$_{0.5}$Ca$_{0.5}$MnO$_{3}$ presented a
maximum at T$_{N}\thickapprox $150 K. For temperatures lower than 10, 20 and
50 K the Nd$_{0.5}$Sr$_{0.5}$MnO$_{3}$, Nd$_{0.5}$Ca$_{0.5}$MnO$_{3}$ and Sm$%
_{0.5}$Ca$_{0.5}$MnO$_{3}$ samples respectively, showed a sharp increase in
the magnetization. This trend have been associated to a short range magnetic
ordering of the intrinsic magnetic moment of Nd$^{3+}$ ions\cite{Mathieu}.
However, no long range ferromagnetic order of the Nd$^{3+}$ ions was found
in neutron diffraction measurements at these low temperatures\cite{Kajimoto}$%
^{,}$ \cite{Millange}.

Differently from the three previous samples, the Dy$_{0.5}$Ca$_{0.5}$MnO$_{3}
$ and Ho$_{0.5}$Ca$_{0.5}$MnO$_{3}$ compounds do not present a strong
magnetization maximum at the charge ordering temperature. However, a clear
inflection is observed at T$_{CO}$ for both samples, as revealed by the
temperature derivative shown in the inset of fig. 1b. The existence of
charge ordering in Dy$_{0.5}$Ca$_{0.5}$MnO$_{3}$ and Ho$_{0.5}$Ca$_{0.5}$MnO$%
_{3}$ was suggested by\ T. Terai et al. \cite{Terai} after studies of
magnetization and resistivity curves. As shown ahead our high temperature
measurements of specific heat present peaks at around the same temperature
interval of the suggested charge ordered transition.

\subsection{Specific heat at high temperatures}

Figure 2 shows specific heat measurements with a zero applied magnetic field
from 2 to 300 K in the Nd$_{0.5}$Sr$_{0.5}$MnO$_{3}$, Nd$_{0.5}$Ca$_{0.5}$MnO%
$_{3}$, Sm$_{0.5}$Ca$_{0.5}$MnO$_{3}$, Dy$_{0.5}$Ca$_{0.5}$MnO$_{3}$ and Ho$%
_{0.5}$Ca$_{0.5}$MnO$_{3}$ samples. In order to facilitate the
visualization, the curves for Nd$_{0.5}$Sr$_{0.5}$MnO$_{3}$ and Ho$_{0.5}$Ca$%
_{0.5}$MnO$_{3}$ were displaced 20 J/mol K upside and downside in figure 2a,
and the curve for Dy$_{0.5}$Ca$_{0.5}$MnO$_{3}$ was displaced 20 J/mol K
downside in figure 2b. Specific heat measurements give information about
both lattice and magnetic excitations. At high temperatures the excitations
from the lattice vibrations are dominant and decrease as the temperature
decreases. The magnetic contribution can be obtained approximately by
subtracting the lattice part from the experimental values.

Continuous lines in figure 2 represent the fitting of the thermal
background, in the interval from 30 to 300 K, by the Einstein model given by:

\begin{equation}
C_{Einstein}=3nR\sum_{i}a_{i}\left[ \frac{x_{i}^{2}\,e^{x_{i}}}{\left(
e^{x_{i}}-1\right) ^{2}}\right]   \label{3}
\end{equation}
where $x_{i}=T_{i}/T$ . We used three optical phonons ($i$ =1, 2, 3) with
energies ${\em T}_{i}$ (in Kelvin) and relative occupations ${\em a}_{i}$.
The Einstein model for the specific heat considers the oscillation frequency
(or energy) independently of the wave vector, which is a valid approximation
for the optical part of the spectrum. The values of temperatures (energies)
and relative occupations are shown in table 1. These values are similar to
those reported, using the same model, by A. P. Ramirez et al.\cite{Ramirez}
in a La$_{0.37}$Ca$_{0.63}$MnO$_{3}$\ sample and Raychaudhuri et al.\cite
{Raychaudhuri} in a Pr$_{0.63}$Ca$_{0.37}$MnO$_{3}$ sample. However, we
should point out that the values found are not unique because there are six
free parameters during the fitting (one energy and one occupation
coefficient for each oscillation mode). Besides, the thermal background
determination is risky, particularly because it could have a ''tail'' of any
magnetic or charge ordering anomaly. Nonetheless, this seems to be the best
possible trial to quantify the specific heat at high temperatures.

Figure 3 represents the differences between the experimental data and the
fitted curves in figure 2. There is a maximum at 231 K for the Nd$_{0.5}$Sr$%
_{0.5}$MnO$_{3}$ sample, which is correlated with the ferromagnetic
transition at 250 K in the corresponding magnetization curve (figure 1). A
second maximum, which could be partially associated to the antiferromagnetic
and charge ordering transitions at 160 K, appears at 180 K. However, lattice
parameters in this compound change rapidly between approximately 110 K and
250 K\cite{Shimomura}. The variation in lattice parameters changes the
intensity of the interactions between the atoms and the oscillation
frequency of the phonons mode, contributing to the specific heat in the
second peak.

Figure 3 also shows that there is a maximum at 243 K for the Nd$_{0.5}$Ca$%
_{0.5}$MnO$_{3}$ sample. This maximum correlates with the corresponding
charge ordering temperature at 250 K. Differently from the Nd$_{0.5}$Sr$%
_{0.5}$MnO$_{3}$ compound, there is no magnetic ordering in this high
temperature interval for the Nd$_{0.5}$Ca$_{0.5}$MnO$_{3}$ sample. However,
in this latter case lattice parameters change very rapidly between
approximately 200 and 250 K \cite{Millange}. An inflection point in the C
vs. T curve (fig. 2a) appears at 141 K for the Nd$_{0.5}$Ca$_{0.5}$MnO$_{3}$
sample. Besides, there is a maximum at 141 K in figure 3, but its height is
relatively small compared to the maximum at 243 K. The Neel temperature
corresponding to this compound is 160 K. Therefore, these results lead us to
conclude that the specific heat variations due to the antiferromagnetic
order are small compared to those induced in the charge ordering and
ferromagnetic transitions.

For the Sm$_{0.5}$Ca$_{0.5}$MnO$_{3}$, Dy$_{0.5}$Ca$_{0.5}$MnO$_{3}$ and Ho$%
_{0.5}$Ca$_{0.5}$MnO$_{3}$ samples, the maxima were found at 254 K, 281 K
and 276 K, respectively. These experimental results and the resistivity
measurements reported by Y. Tokura et al.\cite{Tokura} and T. Terai et al. 
\cite{Terai}, are strong evidences that indicate the existence of CO
transitions in the Sm$_{0.5}$Ca$_{0.5}$MnO$_{3}$, Dy$_{0.5}$Ca$_{0.5}$MnO$%
_{3}$ and Ho$_{0.5}$Ca$_{0.5}$MnO$_{3}$ samples. However, electron
diffraction studies would be needed to unambiguously classify this
transition as CO. The other small peaks, comparable to the experimental
error, were not found to be correlated to any magnetic transition.

Results in figure 3 allow us to calculate the variation in entropy ($\Delta $%
S), within the limits of the model, associated to the charge ordering,
ferromagnetic and antiferromagnetic transitions:

\begin{equation}
\Delta S=\int\limits_{Ti}^{Tf}\frac{\left( C-C_{ph}\right) }{T}\,dT
\label{4}
\end{equation}
where ${\em T}_{i}$ and ${\em T}_{f}$ are two temperatures conveniently
chosen to delimitate the interval of interest and ${\em C}_{ph}$ is the
specific heat due to the lattice oscillations.

For the Nd$_{0.5}$Ca$_{0.5}$MnO$_{3}$ sample the entropy variation between
201 and 301 K, with H=0, was $\Delta S($T$_{CO})=$ 2.0 J/(mol K).
Raychaudhuri et al.\cite{Raychaudhuri} reported an entropy variation, close
to the charge ordering transition in the compound Pr$_{0.63}$Ca$_{0.37}$MnO$%
_{3}$, of 1.8 J/(mol K) with zero applied magnetic field and 1.5 J/(mol K)
with an 8 T magnetic field. On the other hand, Ramirez et al. \cite{Ramirez}
found $\Delta S($T$_{CO})=$ 5 J/(mol K) in a La$_{0.37}$Ca$_{0.63}$MnO$_{3}$
sample. All these results correspond to those expected for a charge ordering
transition \cite{Raychaudhuri}. Besides, the entropy variation (not related
with phonons), calculated between 118 and 201 K for the Nd$_{0.5}$Ca$_{0.5}$%
MnO$_{3}$ sample, was $\Delta S($T$_{N})=$ 0.80 J/(mol K). We associate this
smaller $\Delta {\em S}$\ value to the antiferromagnetic order at T$_{N}=$%
160 K.

The entropy variation between 133 and 274 K for the Nd$_{0.5}$Sr$_{0.5}$MnO$%
_{3}$ sample was $\Delta S($T$_{CO+FM})=$ 3.6 J/(mol K). Considering that
the entropy variation associated to the charge ordering transition is the
same as for the Nd$_{0.5}$Ca$_{0.5}$MnO$_{3}$ sample, one finds that the
entropy variation associated to the ferromagnetic transition is
approximately 1.6 J/(mol K). Using a model proposed by J. E. Gordon et al. 
\cite{Gordon} we estimate that the change in entropy needed to a full
ferromagnetic transition in the Nd$_{0.5}$Sr$_{0.5}$MnO$_{3}$ sample is
12.45 J/(mol K). Therefore, the fact that the entropy variation found is 13
\% of the theoretical value, suggests that only a small part of the spins
order ferromagnetically. J. E. Gordon et al. \cite{Gordon} found also that
the entropy variation associated to the ferromagnetic order in a Nd$_{0.67}$%
Sr$_{0.33}$MnO$_{3}$ sample was approximately 10 \% of the theoretical
value. The entropy variation, not associated to lattice oscillations,
between 200 K and 300 K, for the Sm$_{0.5}$Ca$_{0.5}$MnO$_{3}$ and Dy$_{0.5}$%
Ca$_{0.5}$MnO$_{3}$ samples were $\Delta S($T$_{CO})=$ 1.15 J/(mol K) and
2.80 J/(mol K); between 216 K and 294 K for the Ho$_{0.5}$Ca$_{0.5}$MnO$_{3}$
sample was $\Delta S($T$_{CO})=$ 0.78 J/(mol K). The entropy variation for Sm%
$_{0.5}$Ca$_{0.5}$MnO$_{3}$ and Dy$_{0.5}$Ca$_{0.5}$MnO$_{3}$ are closer to
the ones of the Nd$_{0.5}$Ca$_{0.5}$MnO$_{3}$ sample. The $\Delta S($T$_{CO})
$\ for the Ho$_{0.5}$Ca$_{0.5}$MnO$_{3}$\ sample is smaller than for the
other samples, suggesting a different nature of the charge ordering
transition.

Figure 4 shows the specific heat measurements, between 200 and 300 K, with a
zero applied magnetic field (open symbols) and with H = 9 T (closed symbols)
for the five measured samples. To allow comparisons the Nd$_{0.5}$Ca$_{0.5}$%
MnO$_{3}$ curve was displaced 20 J/mol K upside; the Sm$_{0.5}$Ca$_{0.5}$MnO$%
_{3}$, Dy$_{0.5}$Ca$_{0.5}$MnO$_{3}$ and Ho$_{0.5}$Ca$_{0.5}$MnO$_{3}$
curves\ were displaced 20, 40 and 60 J/mol K downside, respectively. The
external magnetic field suppresses the peak around the ferromagnetic
temperature in the Nd$_{0.5}$Sr$_{0.5}$MnO$_{3}$ sample. Although a 9 T
magnetic field is much smaller than a thermodynamic field given by {\em k}$%
_{B}${\em T}$_{{\em C}}${\em \ / }$\mu _{B}$, near the ferromagnetic phase
transition the system is in an unstable condition, which is very sensitive
to the external magnetic field. This fact explains the suppression of the
peak in the presence of field in the specific heat curve and the decrease in
the slope of the magnetization curve (not shown).

On the other hand, the external magnetic field seems to have no effect in
the specific heat of the other four compounds. This result indicates that,
although the sample magnetization increases with temperature close to the
charge ordering transition, the CO phase do not appear to depend
thermodynamically on the application of an external magnetic field. The
increase of magnetization close to T$_{CO}$ is associated with an abrupt
change in lattice parameters\cite{Tokura}, affecting the distance among
magnetic ions and hence their interactions. Our experiments show that an
external magnetic field of 9 T is not strong enough to produce variations in
the specific heat of the CO phase occurring in these samples.

\section{Conclusions}

We have made a magnetic characterization of Nd$_{0.5}$Sr$_{0.5}$MnO$_{3}$, Nd%
$_{0.5}$Ca$_{0.5}$MnO$_{3}$, Sm$_{0.5}$Ca$_{0.5}$MnO$_{3}$, Dy$_{0.5}$Ca$%
_{0.5}$MnO$_{3}$ and Ho$_{0.5}$Ca$_{0.5}$MnO$_{3}$ polycrystalline samples.
Ferromagnetic, antiferromagnetic and charge ordering transitions in our
samples agreed with previous reports. We also reported, to our knowledge for
the first time, specific heat measurements with applied magnetic fields
between 0 and 9 T and temperatures between 2 and 300 K in all these five
samples. Each curve was successfully fitted at high temperatures by an
Einstein model with three optical phonon modes. Close to the charge ordering
and ferromagnetic transition temperatures the specific heat curves showed
peaks superposed to the characteristic response of the lattice oscillations.
The entropy variation corresponding to the charge ordering transition was
higher than the one corresponding to the ferromagnetic transition. Near 160
K the specific heat curve showed an abrupt change in slope for the two
compounds with Nd$^{3+}$ ions, which were correlated to the corresponding
antiferromagnetic transition. However, an external 9 T magnetic field seems
to have no effect in the specific heat of the CO phase transition.

We thank Dr. P. N. Lisboa-Filho for the preparation of Nd$_{0.5}$Sr$_{0.5}$%
MnO$_{3}$, Nd$_{0.5}$Ca$_{0.5}$MnO$_{3}$ and Ho$_{0.5}$Ca$_{0.5}$MnO$_{3}$
samples and the Brazilian science agencies FAPESP and CNPq for the financial
support.

\bigskip

\newpage

\begin{table}[tbp]
\caption{Values of energies ${\em T}_{i}$ (in Kelvin) and relative
occupations ${\em a}_{i}$ for the three optical phonons ($i$ =1, 2, 3) in an
Einstein model for the specific heat in all the studied samples.}
\bigskip
\par
\begin{tabular}{ccccccc}
Sample & T$_{1}$(K) & T$_{2}$(K) & T$_{3}$(K) & a$_{1}$ & a$_{2}$ & a$_{3}$
\\ 
Nd$_{0.5}$Sr$_{0.5}$MnO$_{3}$ & 148 & 438 & 997 & 0.30 & 0.64 & 0.11 \\ 
Nd$_{0.5}$Ca$_{0.5}$MnO$_{3}$ & 152 & 432 & 1035 & 0.27 & 0.62 & 0.18 \\ 
Sm$_{0.5}$Ca$_{0.5}$MnO$_{3}$ & 157 & 450 & 846 & 0.27 & 0.64 & 0.16 \\ 
Dy$_{0.5}$Ca$_{0.5}$MnO$_{3}$ & 126 & 351 & 821 & 0.17 & 0.41 & 0.51 \\ 
Ho$_{0.5}$Ca$_{0.5}$MnO$_{3}$ & 147 & 438 & 1023 & 0.27 & 0.51 & 0.25
\end{tabular}
\end{table}
\begin{figure}[tbp]
\caption{Temperature dependence of the magnetization, with a 5 T applied
magnetic field, in field cooling--warming condition for the five
polycrystalline samples studied. Magnetization is given in Bohr magnetons
per manganese ion. The Curie (T$_{C}$), Ne\'{e}l (T$_{N}$) and charge
ordering (T$_{CO}$) temperatures are indicated for each curve. The curves
are plotted with a logarithmic scale in the y-axes to allow the comparison
of all samples. The inset in fig.1b represents the temperature derivative of
the magnetization close to the charge ordering transition.}
\label{Figure 1}
\end{figure}

\begin{figure}[tbp]
\caption{Specific heat measurements between 2 and 300 K in the five measured
samples. Continuous lines represent the fitting of the phonon background to
the Einstein model (see text). The Nd$_{0.5}$Sr$_{0.5}$MnO$_{3}$ and Ho$%
_{0.5}$Ca$_{0.5}$MnO$_{3}$ curves in fig.2a were displaced 20 J/mol K upside
and downside, respectively; the Ho$_{0.5}$Ca$_{0.5}$MnO$_{3}$ curve in
fig.2b was displaced 20 J/mol K downside.}
\label{Figure 2}
\end{figure}

\begin{figure}[tbp]
\caption{Differences between the experimental specific heat data and the
corresponding background curves in Nd$_{0.5}$Sr$_{0.5}$MnO$_{3}$, Nd$_{0.5}$%
Ca$_{0.5}$MnO$_{3}$, Sm$_{0.5}$Ca$_{0.5}$MnO$_{3}$, Dy$_{0.5}$Ca$_{0.5}$MnO$%
_{3}$ and Ho$_{0.5}$Ca$_{0.5}$MnO$_{3}$ samples.}
\label{Figure 3}
\end{figure}

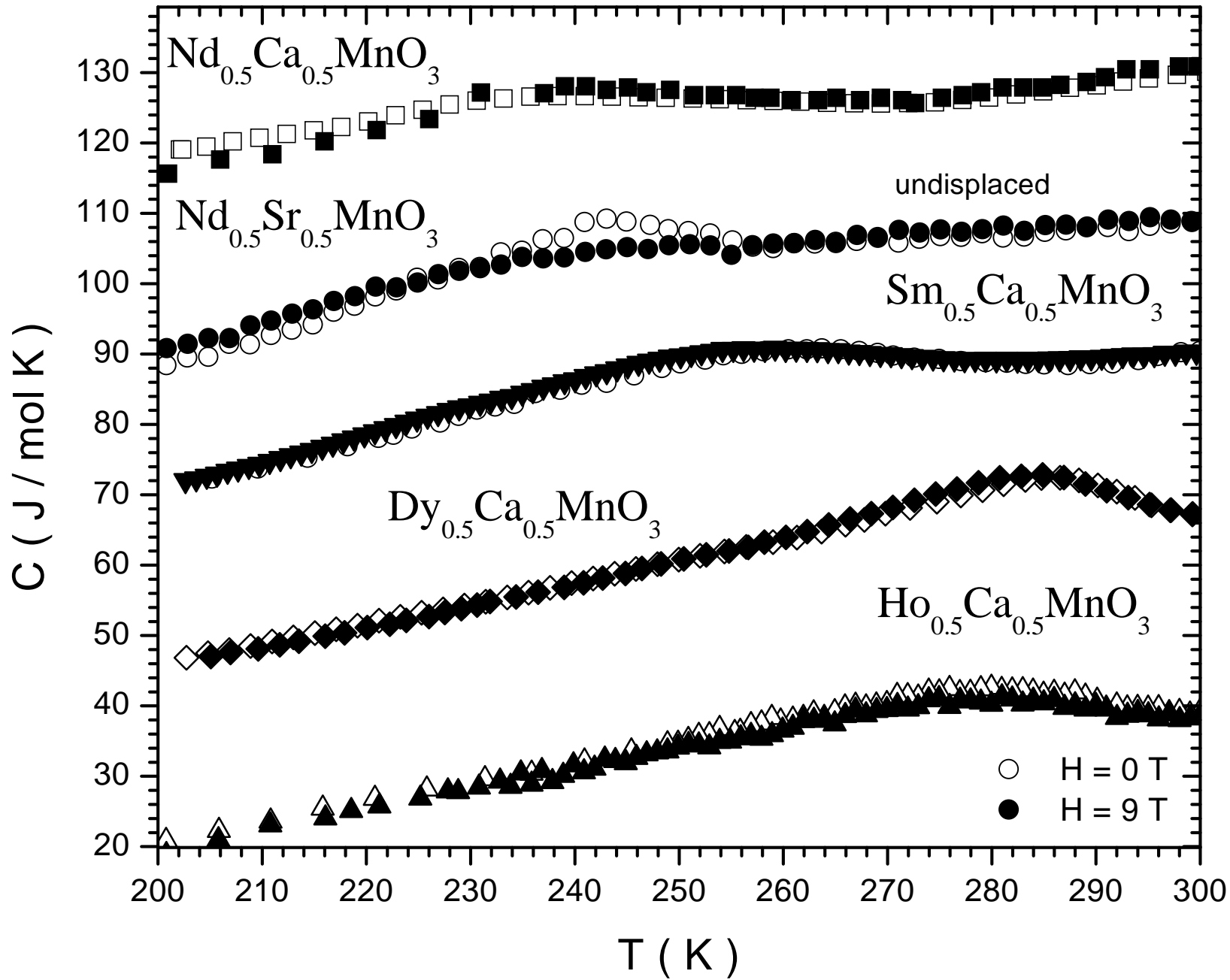
\begin{figure}[tbp]
\caption{Specific heat measurements between 200 and 300 K with a zero
applied magnetic field (open symbols) and with H = 9 T (closed symbols) in
the five measured samples. To allow comparisons the Nd$_{0.5}$Ca$_{0.5}$MnO$%
_{3}$ curves were displaced 20 J/mol K upside; the Sm$_{0.5}$Ca$_{0.5}$MnO$%
_{3}$, Dy$_{0.5}$Ca$_{0.5}$MnO$_{3}$ and Ho$_{0.5}$Ca$_{0.5}$MnO$_{3}$
curves\ were displaced 20, 40 and 60 J/mol K downside, respectively.}
\label{Figure 4}
\end{figure}

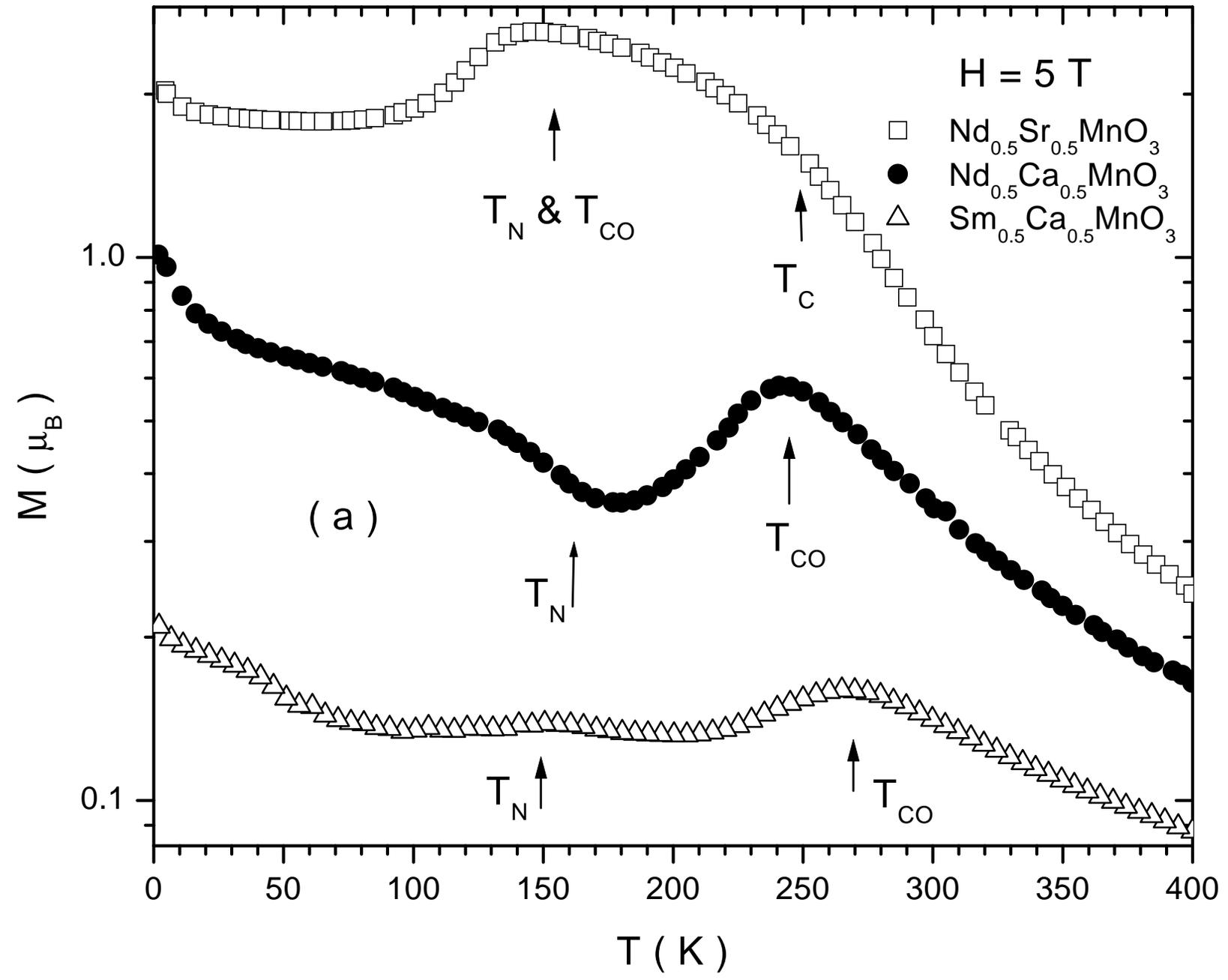
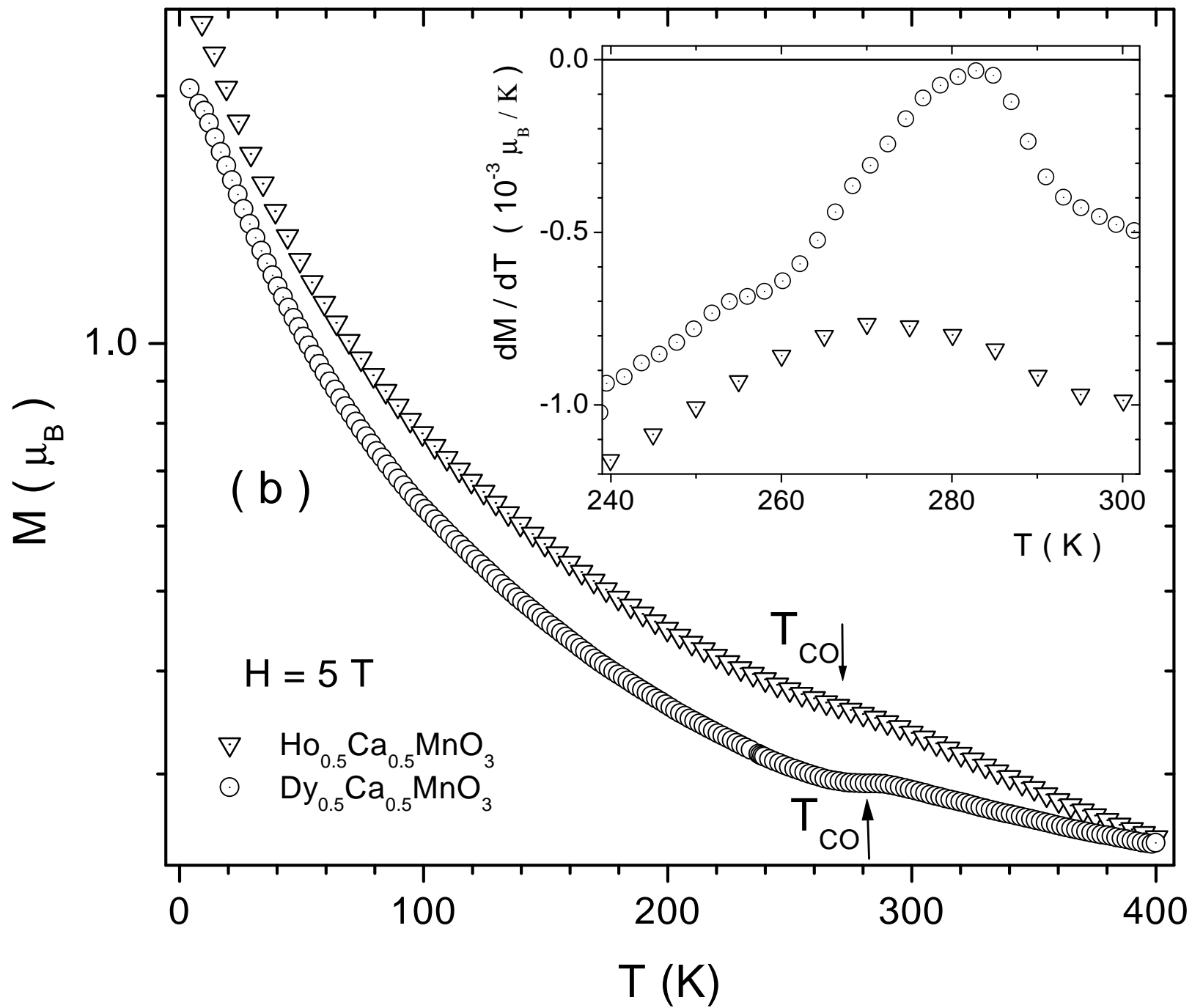
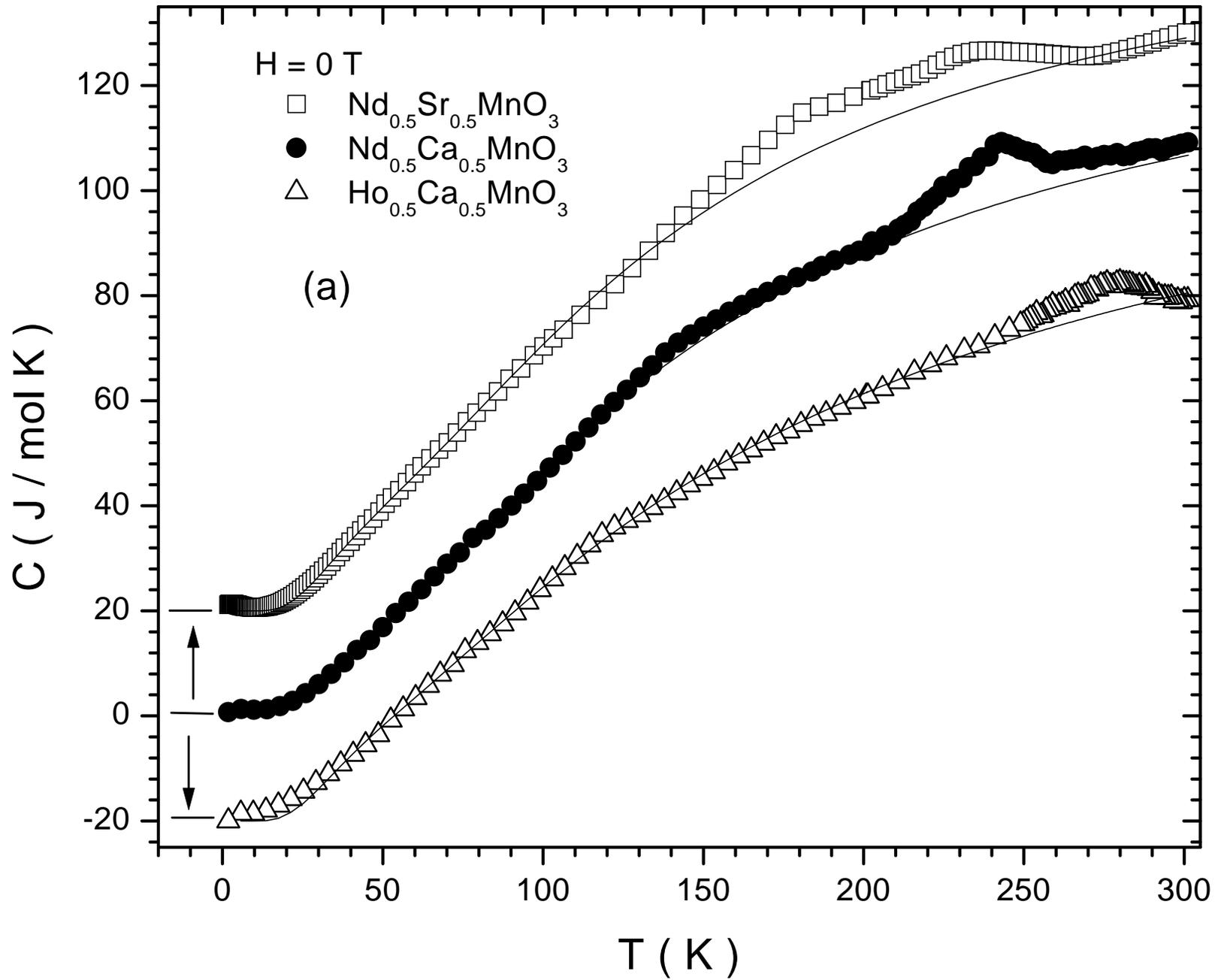
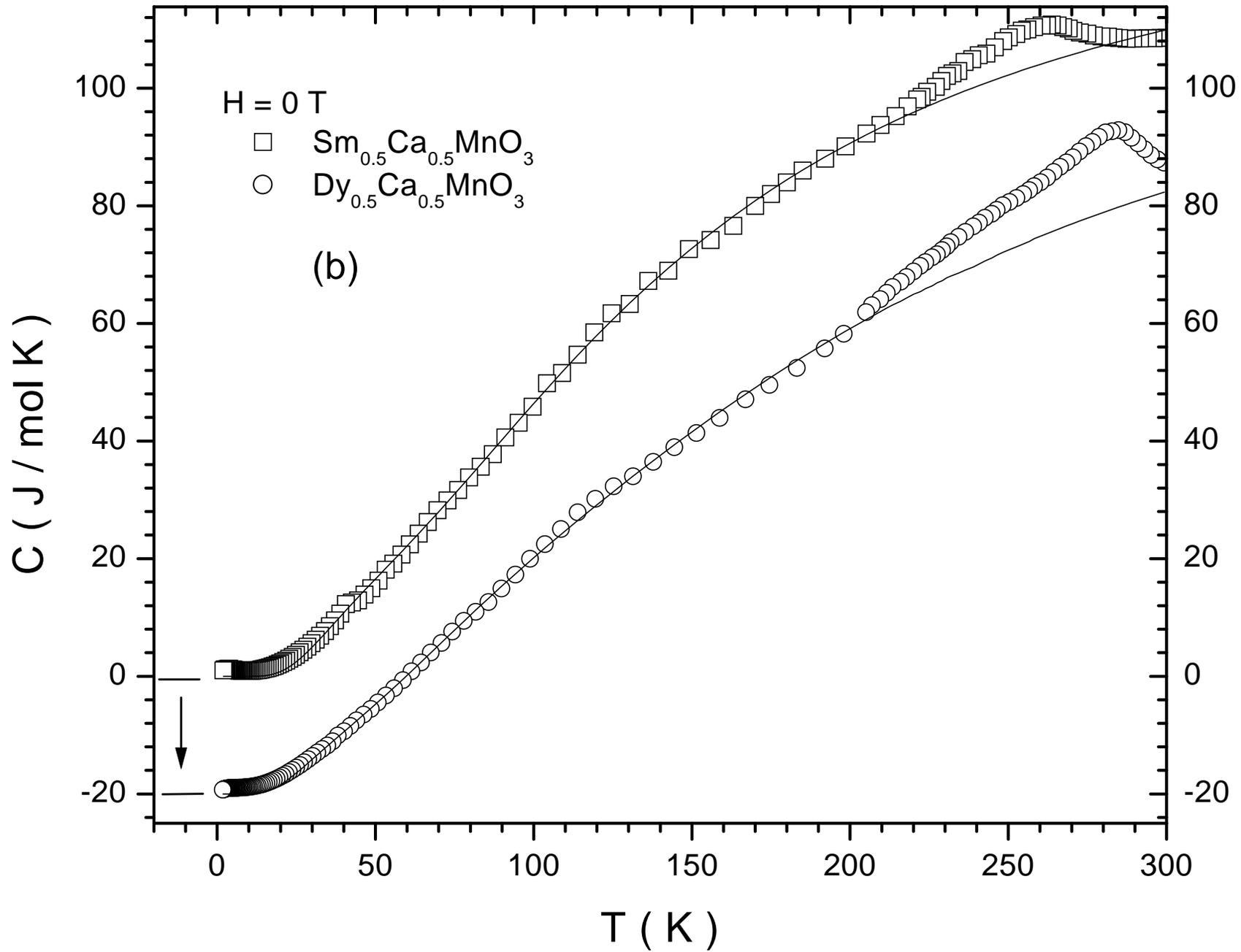
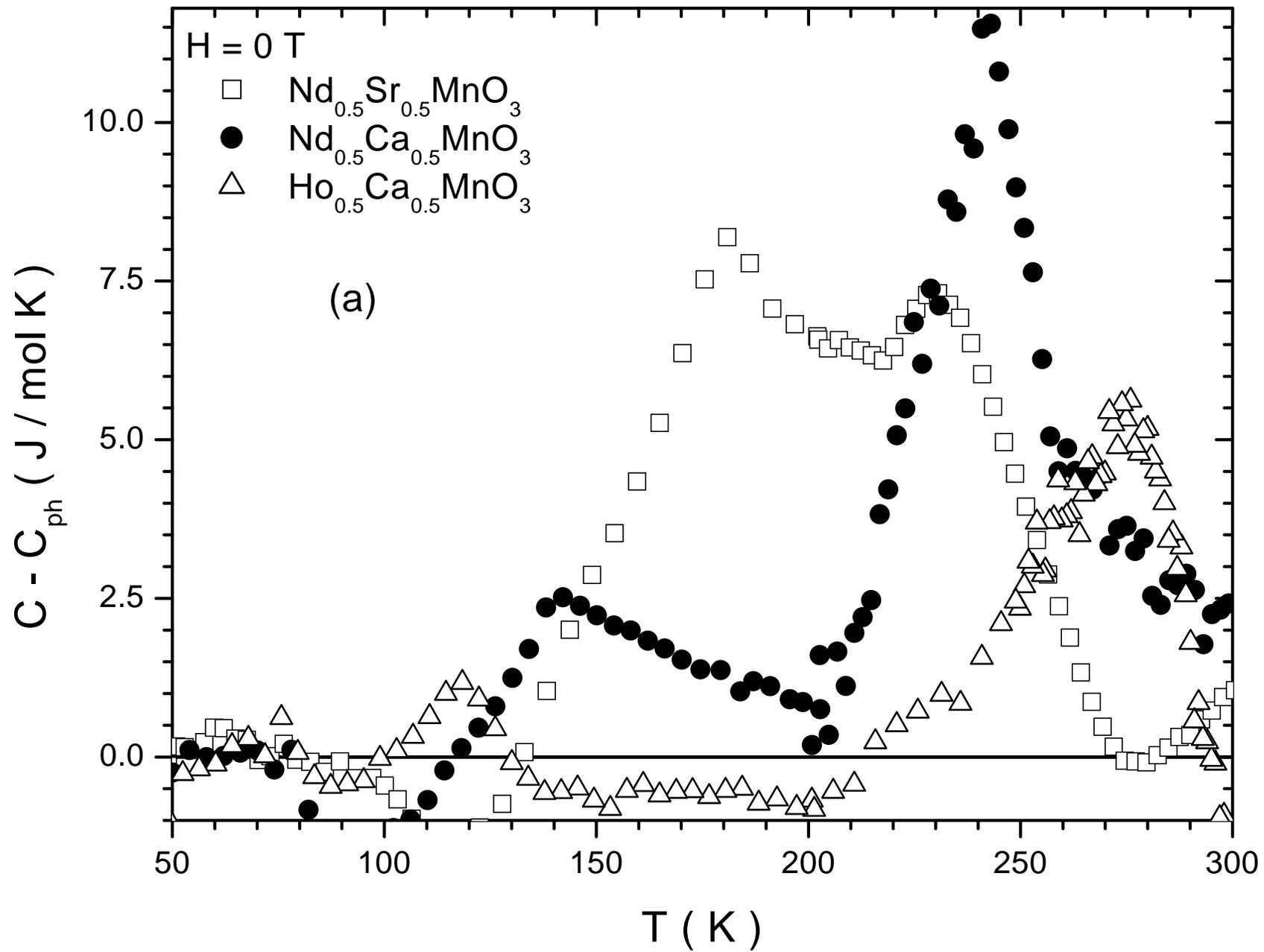
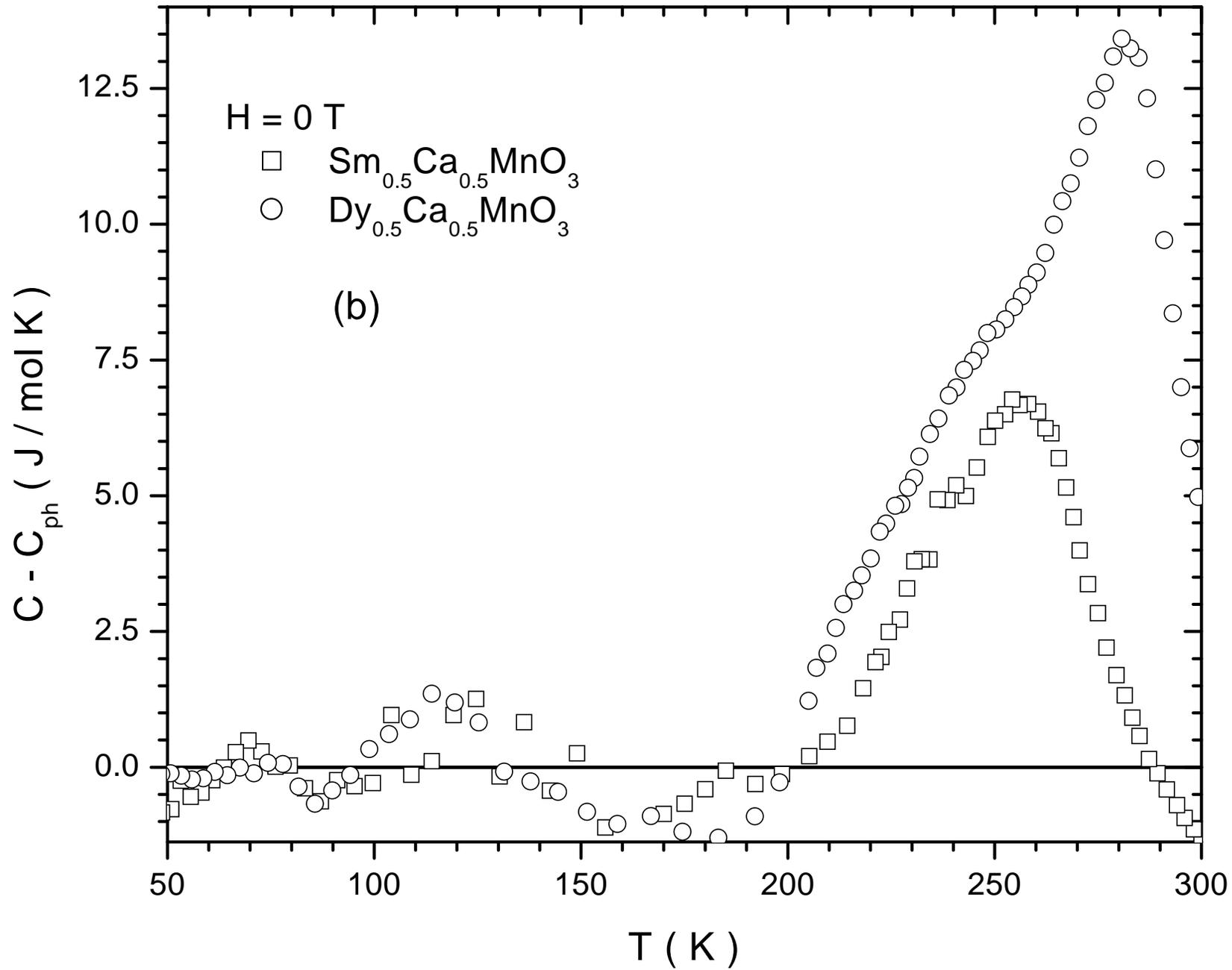
\end{document}